\documentclass[aps,prd,twocolumn,superscriptaddress,footinbib]{revtex4-2}
\pdfoutput=1
\usepackage{graphicx}
\usepackage{subcaption}
\graphicspath{{./picture/}}
\usepackage{empheq}
\usepackage{amssymb}
\usepackage{mathtools}
\usepackage{commath}
\usepackage{verbatim}
\usepackage{xcolor}
\usepackage{soul}
\usepackage{booktabs}
\usepackage{amsmath}
\usepackage{hhline}
\usepackage{url}
\usepackage{ragged2e}
\usepackage{caption}
\DeclareCaptionFormat{revtex}{#1#2\justifying{#3}}
\usepackage{epsfig}
\usepackage{mathrsfs}
\usepackage{nicefrac} 
\usepackage{upgreek}
\usepackage{bbm}
\usepackage{psfrag}

\usepackage{hyperref}% add hypertext capabilities
\newcommand{\mpl}{M_\mathrm{pl}}

\usepackage[inline,shortlabels]{enumitem}

\newcommand{\g}{\mathsf{g}}

\newcommand{\V}{\mathcal{V}}
\newcommand{\Vend}{\mathcal{V}_{\rm end}}
\newcommand{\x}{\textbf{x}}

\newcommand{\wrehbar}{\bar{w}_{\rm re}}

\newcommand{\Nreh}{N_{\rm re}}
\newcommand{\X}{X}

\newcommand{\M}{M}
\usepackage{slashed}				%Feynman slash notation
\newcommand{\Treh}{T_{\rm re}}
\newcommand{\GG}{\Gamma}
\newcommand{\Nk}{N_{\rm k}}

\newcommand{\sigmar}{\sigma_r}
\newcommand{\sigmans}{\sigma_{n_s}}
\newcommand{\nsbar}{\bar{n}_s}
\newcommand{\rbar}{\bar{r}}

\newcommand{\lambdaphi }{\lambda}

\newcommand{\modified}[1]{\textcolor{black}{#1}}

\begin{document}

\title{Observing Cosmic Reheating with the expanded Simons Observatory}

\author{Lei Ming}
\email{minglei@scnu.edu.cn}
\affiliation{Key Laboratory of Atomic and Subatomic Structure and Quantum Control (Ministry of Education), Guangdong Basic Research Center of Excellence for Structure and Fundamental Interactions of Matter, School of Physics, South China Normal University, Guangzhou 510006, China} 
\affiliation {Guangdong Provincial Key Laboratory of Quantum Engineering and Quantum Materials, Guangdong-Hong Kong Joint Laboratory of Quantum Matter, South China Normal University, Guangzhou 510006, China}

\author{Marco Drewes}
\email{marco.drewes@uclouvain.be}
\affiliation{\modified{UCLouvain, Centre for Cosmology, Particle Physics and Phenomenology (CP3), Chemin du Cyclotron 2, 1348 Louvain-la-Neuve, Belgium}}
\affiliation{Physik–Department, Technische Universität München, James Franck Straße 1, D-85748 Garching, Germany}

%\date{\today}

\begin{abstract}
The Simons Observatory will be extended by three Small Aperture Telescopes by 2027, increasing the total number of these instruments to six. We study the prospects for probing the reheating temperature and the inflaton coupling with this configuration, assuming a discovery of primordial gravitational waves in benchmark scenarios with a tensor-to-scalar ratio $r=0.0036$ or $r=0.01$. In popular plateau models of inflation, such an observation would fix the scale of inflation and enable determination of the order of magnitude of the reheating temperature and the inflaton interactions. For QCD-driven Warm Inflation the reheating temperature and inflaton coupling to gluons could, under optimistic assumptions, be measured with a precision of a few percent. Such a measurement would imply a clear prediction for complementary inflaton searches in axion experiments, paving the way toward probing the mechanism responsible for the initial conditions of the hot Big Bang in the laboratory.
\end{abstract}

\maketitle

\begin{footnotesize}
%\tableofcontents
\end{footnotesize}
\section{Introduction}

The theory of a Hot Big Bang \cite{Friedman:1922kd,Lemaitre:1927zz,Alpher:1948ve} has become the standard paradigm in modern cosmology.
More precisely, the so-called concordance model of cosmology -- or $\Lambda$CDM model -- can explain the vast set of observational data \cite{Cortes:2026yde} with a very small number of six parameters \footnote{See \cite{CosmoVerseNetwork:2025alb} for an overview of hints for deviations from the $\Lambda$CDM model.}. 
However, while observations prove that the established fundamental laws of Nature \cite{ParticleDataGroup:2024cfk}, i.e., the Standard Model of particle physics (SM) and General Relativity hold in the most distant regions of the observable Universe \footnote{Established observational facts that cannot be described in this framework include the microphysical composition of the Dark Matter \cite{Cirelli:2024ssz} and
the origin of the matter-antimatter asymmetry of the Universe \cite{Canetti:2012zc}; it is not clear whether the explanations for other tensions or anomalies \cite{Abdalla:2022yfr,CosmoVerseNetwork:2025alb} require the existence of New Physics.}, it remains unclear what mechanism set the initial conditions of the hot big bang -- including the overall homogeneity and isotropy of the cosmos, its small overall spacial curvature, the origin of the tiny temperature fluctuations that formed the seeds for galaxy formation, and the initial temperature $\Treh$ at the onset of the radiation dominated epoch.

The leading candidate for an explanation of these initial conditions is provided by \emph{cosmic inflation} \cite{Starobinsky:1980te,Guth:1980zm,Linde:1981mu}, i.e., the idea that the early Universe underwent a phase of accelerated cosmic expansion. However, it remains unclear what mechanism drove the acceleration, and how it can be implemented into a fundamental theory of Nature beyond the SM. 
Many of the proposed models can be parameterised in terms of a single scalar field $\varphi$, dubbed inflaton, with a potential $\V(\varphi)$~\cite{Martin:2013tda}. Unfortunately current observational data are insufficient to single out a particular choice of $\V$ \cite{Martin:2024qnn}.
Moreover, even less is known about the interactions between $\varphi$ and other fields connecting inflation to theories of particle physics and the SM. 

 \emph{Cosmic reheating}~\cite{Albrecht:1982mp,Dolgov:1989us,Traschen:1990sw,Shtanov:1994ce,Kofman:1994rk,Boyanovsky:1996sq,Kofman:1997yn}, i.e., the dissipative transfer of energy from $\varphi$ to other degrees of freedom that heated the universe after inflation,
provides an important probe of this connection.
It is well-known that the reheating epoch 
leaves an imprint in the Cosmic Microwave Background (CMB) \cite{Lidsey:1995np}, from which one can obtain information on $\Treh$ \cite{Martin:2010kz,Adshead:2010mc,Mielczarek:2010ag,Easther:2011yq,Dai:2014jja} and the microphysics of reheating \cite{Drewes:2015coa,Martin:2016iqo}.
While present data already contain some information on reheating~\cite{Martin:2014nya,Martin:2016oyk}, 
upcoming missions could for the first time provide a measurement \footnote{By this we mean the ability to impose both an upper and a lower bound on a quantity from data.} 
of $\Treh$ \cite{Martin:2014rqa} and the underlying microphysics~\cite{Drewes:2019rxn}. 
An important step forward in this direction would be the discovery of primordial gravitational waves.
After the funding for CMB-S4 \cite{CMB-S4:2020lpa} has been halted, the two most sensitive experiments that can achieve this goal 
are the Japanese LiteBIRD satellite \cite{Sugai:2020pjw} and the Simons Observatory (SO) \cite{SimonsObservatory:2018koc}.
The Chinese  AliCPT array \cite{Li:2017drr,Li:2018rwc}
will provide data complementary to the SO due to its geographic location in the Northern Hemisphere.
We have previously assessed the sensitivity of LiteBIRD \cite{Drewes:2022nhu,Drewes:2023bbs} and 
AliCPT-1 \cite{Liu:2025sut} to reheating. 
In the present work we study the perspectives to probe reheating and the connection between inflation and particle physics with the expanded SO \cite{SimonsObservatory:2025wwn, SimonsObservatory:2025avm}.

Currently three small aperture telescopes (SATs) are deployed on the SO site include, two mid-frequency (MF) SATs (operating at 93 and 145 GHz) and one ultra-high-frequency SAT operating at 225
and 280 GHz. 
This number will be increased to six, including two more MF SATs from the UK and one low-frequency (LF) SAT from Japan, the latter operating at 27 and 39 GHz. This will considerably improve the SO sensitivity to CMB B-modes produced by primordial gravitational waves from inflation \cite{SimonsObservatory:2025avm}. 
Our primary goal is to quantify what level of sensitivity to $\Treh$ and the microphysical inflaton coupling $\g$ that heated the universe can be achieved by this. We consider four models of inflation -- 
namely 
$\alpha$-attractor T-model ($\alpha$-T)~\cite{Kallosh:2013maa,Kallosh:2013hoa,Carrasco:2015pla,Carrasco:2015rva},
radion gauge inflation (RGI)~\cite{Fairbairn:2003yx,Martin:2013nzq}, mutated hilltop inflation (MHI)~\cite{Pal:2009sd,Pal:2017bmd}  and 
QCD-driven warm inflation (QCD-WI) \cite{Berghaus:2025dqi}
-- and two fiducial values $\rbar= 0.0036$ and $\rbar = 0.01$ for the tensor-to-scalar ratio $r$ in the observational benchmark scenarios summarized in table \ref{PrincipalBenchmark}.

\section{Models}
For the purpose of our analysis we can express the information on reheating contained in the CMB in terms of three numbers, the amplitude and spectral index of scalar perturbations $A_s$ and $n_s$, respectively, as well as $r$. 
A fundamental problem that limits the testability of the inflationary paradigm is that the number of unknown model parameters generally exceeds the number of the three observables $\{A_s, n_s, r\}$. 
Any given set of observed values of $\{A_s, n_s, r\}$ can be explained with a large number of inflationary models, defined by the choice of $\V$. 
Moreover, even within a given model, it is often not possible to constrain all parameters from observation. Many popular models contain several unknown parameters in $\V$ \footnote{Notable exceptions are Starobinsky inflation \cite{Starobinsky:1980te}, Higgs inflation \cite{Bezrukov:2007ep} and the QCD-WI model \eqref{eq:lag} used here.}. Reheating adds at least one parameter; for fixed $\V$ we can choose this to be either the duration of reheating in terms of $e$-folds $\Nreh$ or $\Treh$, cf.~appendix \ref{TrefromCMB}. While it is in principle possible to measure $\Treh$ along with two parameters in $\V$ from three observables, in practice the error bars are usually too large to obtain a meaningful constraint \cite{Drewes:2023bbs}.

The problem becomes far worse at the level of fundamental (microphysical) parameters: For given $\V$, $\Treh$ is determined by the efficiency of particle production during reheating, which can be expressed in terms of a friction coefficient $\GG$. Since the reheating process is generally non-linear and can be strongly affected by feedback from the produced particles \cite{Amin:2014eta}, $\GG$ can depend on a large set of unknown microphysical parameters (related to the particles' properties and their interactions) that affect the reheating process.
Hence, even within a given model of inflation the values of $\{A_s, n_s, r\}$ depend on a potentially large set of microphysical parameters, and measuring these observables usually does not lead to an independent prediction that can be tested empirically.

One of the rare models where such a prediction can be made is QCD-WI  \cite{Berghaus:2025dqi}, an extension of the SM by a pseudoscalar field $\varphi$ that couples to gluons,
\begin{equation}
\label{eq:lag}
\mathcal{L} = \mathcal{L}_{\rm SM} + \frac{1}{2} \partial^\mu \varphi \partial_\mu \varphi 
 - \frac{\upalpha_s}{8 \pi}\frac{\varphi}{f} G^{a}_{\mu \nu} \tilde{G}^{a\mu \nu}
%- \lambda \varphi^4
- \V(\varphi) ,
\end{equation}
with $\upalpha_s$ the coupling of quantum chromodynamics (QCD), $G^{a}_{\mu \nu}$ the gluon field strength tensor and $f$ a decay constant. 
The friction coefficient is dominated by sphaleron heating \cite{McLerran:1990de,Laine:2021ego} and reads 
$\GG  \simeq N_c^5 \alpha^5_s \frac{T^3}{2f^2}$
\footnote{In the regime where $\GG$ strongly affects the dynamics of the background field $\varphi$ a better approximation is given by	$\GG  = 
	N_c^5 \alpha^5_s T^3 /[2f^2 (1 + 2 N_f N_c^4 \alpha^5_s T\mpl/ 
    (\V/3)^{1/2}
    )]$ \cite{Berghaus:2025dqi}.}  with $N_c=3$, 
    It is sufficient to drive an extended period of warm inflation \cite{Berera:1995ie,Kamali:2023lzq}. 
A key difference to the QCD axion \cite{Peccei:1977ur,Peccei:1977hh,Wilczek:1977pj,Weinberg:1977ma} lies in the slight breaking of the shift symmetry \footnote{This breaking is sufficiently soft that $\V$ is protected from the problems \cite{Yokoyama:1998ju} which haunt the original warm inflation proposal \cite{Berera:1995ie}. Specifying its origin in the UV is not required for the present analysis; it could, e.g., originate from gravitational effects \cite{Karananas:2025ews}.}
by the potential needed for the slow rolling during inflation, for which we pick $\V=\lambda \varphi^4$. Nevertheless, from an experimental viewpoint the inflaton $\varphi$ behaves like a QCD-axion and can be found in axion search experiments \cite{Antel:2023hkf}. Since a discovery of primordial tensor modes in the CMB can be translated into a  measurement of the decay constant $f$, the model \eqref{eq:lag} is truly testable in the sense that a measurement of $\{A_s, n_s, r\}$ would enable a prediction that can potentially be verified in the laboratory, providing a unique opportunity to probe the mechanism that set the stage for the hot big bang and its connection to fundamental physics experimentally \footnote{In addition to finding the inflaton in the laboratory, observing the cosmological bispectrum \cite{Mirbabayi:2022cbt} would provide another test of QCD-WI.}.

In addition to the benchmark model \eqref{eq:lag} we also consider three more conventional (cold) inflation models, namely the $\alpha$-T, RGI and MHI models 
with the potentials 
\begin{eqnarray}
	{\rm[}\alpha{\rm -T]} \quad \V&=&\M^4{\rm tanh}^{2
		%\nn
	}
	\left(\frac{\varphi}{\sqrt{6\alpha}\mpl}\right) 
	\label{alpha V} \\
    {\rm[RGI]} \quad \V &=& \M^4\frac{(\varphi/M_{pl})^2}{\alpha+(\varphi/\mpl)^2}.\label{RGI V} \\
    {\rm[MHI]} \quad \V &=& \M^4\left[1-\frac{1}{{\rm cosh}(\varphi/(\alpha \mpl))}\right]~. \label{MHI V} 
\end{eqnarray}   
Each of these containts two parametets in $\V$, the scale $\M$ of the inflaton potential and a parameter $\alpha$ that controls the ratio $m_\phi/\M$, with $m_\phi$ the inflaton mass. 
The latter is defined by the expansion
\begin{eqnarray}\label{Taylor}
 \V = 
 %\sum_\n\frac{\vv_{\n}}{\n !} \frac{\varphi^\n}{\EFTscale^{\n-4}}  =
 \frac{1}{2}m_\varphi^2\varphi^2+\frac{g_\phi}{3!}\varphi^3+\frac{\lambdaphi }{4!}\varphi^4 + \ldots. 
 \end{eqnarray}
Together with $\Treh$, a choice of $\alpha$ and $\M$ can uniquely predict the values of $\{A_s, n_s, r\}$.
At the level of microphysical parameters $\Treh$ depends on  at least one coupling constant $\g$ between $\varphi$ and other fields (unless reheating proceeds entirely through gravitational interactions).
In the models \eqref{alpha V}, \eqref{RGI V} and \eqref{MHI V} bounds on $\Treh$ can in general only be unambiguously translated into information on the microphysical coupling constant $\g$ if \cite{Drewes:2019rxn}
\begin{eqnarray}\label{PerturbativityContraintsPleteau}
\lambdaphi 
\ll
3\pi^2  r A_s 
\ , \  
%and 
|\g| \ll \left(3\pi^2  r A_s\right)^{1/2}.
\end{eqnarray}
For a Yukawa coupling to fermions 
$y\bar{\psi}\psi$ this range can be considerably extended under mild model assumptions, see appendix B in \cite{Liu:2025sut}.
 In the following we parametrize $\GG = y^2 m_\phi / (8\pi)$ and relate $\Treh$ to $y$ with the help of the relations \eqref{Tre} and \eqref{GammaConstraint} in the appendix \footnote{As a parametrization of the duration of reheating, the use of $y$ does not introduce any further assumptions and can (for fixed $\V$ and $g_*$) be used equivalently to $\Treh$ or $\Nreh$, only its interpretation as a microphysical coupling constant is ambiguous if \eqref{PerturbativityContraintsPleteau} is strongly violated \cite{Drewes:2019rxn,Drewes:2026uor}.}.

\section{Method}

Our goal is to constrain a set of model parameters $\X$ from observational data $\mathcal{D}$. 
Fundamentally the model \eqref{eq:lag} contains two dimensionless parameters $f/\mpl$ and $\lambda\}$, while predicting CMB spectra in the models 
\eqref{alpha V} -- \eqref{MHI V}
requires three parameters $\M/\mpl, \alpha, y$ each.
For practical purposes it is preferable to use the logarithmic variables $\x=\log_{10}(f/{\rm GeV})$ for QCD-WI and   $\x=\log_{10}y$ for the plateau models \eqref{alpha V}-\eqref{MHI V}, and to measure all units in GeV. 
Fixing $A_s = 10^{-10}e^{3.043}$ \cite{Planck:2018vyg} 
does not significantly impact the sensitivity forecast \cite{Drewes:2023bbs} and
allows to eliminate one of the model parameters \footnote{For the model \eqref{eq:lag} the constraint imposed by fixing $A_s$ can only be implemented numerically, for the cold plateau models the analytic relations are e.g.~given in \cite{Drewes:2023bbs}.}, which we choose to be $\lambda$ for QCD-WI and $\M$ in the plateau models, leaving us with $\X=\x$ and $\X = \{\alpha,\x\}$ in the former and latter, respectively.
The relations between the parameters $\X$ and the observables $\{A_s,n_s,r\}$  are known for the models considered here, we provide a brief summary in Appendix \ref{TrefromCMB}.

We quantify the knowledge gain on the parameters $\X$ from data $\mathcal{D}$ collected in SO observations in terms of one- and two-dimensional posterior probability distributions.
We employ the simple analytic method introduced in \cite{Drewes:2022nhu}, which has been tested against full MCMC forecasts in \cite{Drewes:2023bbs}. 
With $A_s$ fixed, the relevant data can be parameterised as $\mathcal{D} = \{n_s,r\}$. 
The foreseen sensitivity for given fiducial values $\nsbar$ and $\rbar$
can in good approximation be described by 
a likelihood function \begin{eqnarray}\label{Eq:Likelihood}
	P(\mathcal{D}|\X) = C_2\mathcal{N}(n_s,r|\nsbar,\sigmans;\rbar,\sigmar)\theta(r) 
\end{eqnarray}
with $\mathcal{N}(n_s,r|\nsbar,\sigmans;\rbar,\sigmar)$
a two-dimensional Gaussian  with variances $\sigmans$ and $\sigmar$ for given fiducial values $r=\rbar$ and $n_s=\nsbar$. 
The off-diagonal correlations are negligible \cite{Drewes:2023bbs}, which is related to the fact that $r$ is primarily determined by the SATs while $n_s$ is measured with large aperture telescopes. 
We can then quantify the knowledge gain about the set of model parameters $\X$ from data $\mathcal{D} = \{n_s,r\}$ in terms of a posterior distribution $P(\X|\mathcal{D})=P(\mathcal{D}|\X)P(\X)/P(\mathcal{D})$, where 
\begin{align}
    P(\mathcal{D})=\int d\X P(\mathcal{D}|\X)P(\X).
\end{align}
Demanding that $\int P(\mathcal{D}|\X)d\mathcal{D} = 1$ fixes the constant $C_2$. 
We use a flat prior $P(\X) = \theta(\Nreh)\theta(\Treh - T_{\rm BBN})$, with $T_{\rm BBN} = 10$ MeV to assure successful big bang nucleosynthesis (BBN), see e.g.~\cite{Barbieri:2025moq}.

\begin{figure}
	\centering
    \includegraphics[width=0.9\linewidth]{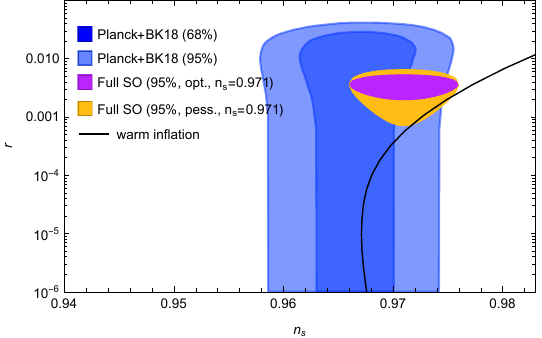}\\ \includegraphics[width=0.9\linewidth]{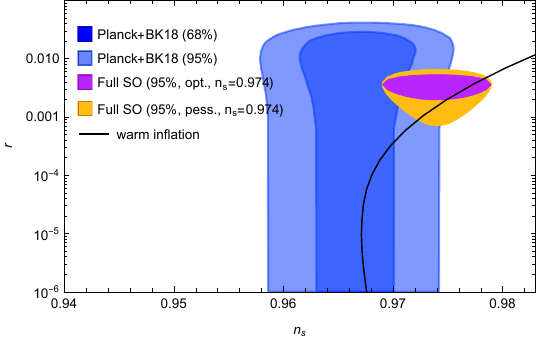}
	\caption{\emph{Upper panel:} 95\% confidence region of the likelihood function \ref{Eq:Likelihood} 
    for benchmarks \ref{it:pessC} (yellow) and
    \ref{it:optC} (violet)
    compared to the prediction from the QCD-WI model \eqref{eq:lag} (black line) and current observational constraints from Planck+BK (blue). 
    \emph{Lower panel}:
    Same for benchmarks \ref{it:pessD} (yellow) and
    \ref{it:optD} (violet).
    } 
	\label{fig:ellipsesWI}
\end{figure}

\begin{figure}
	\centering
    \includegraphics[width=0.9\linewidth]{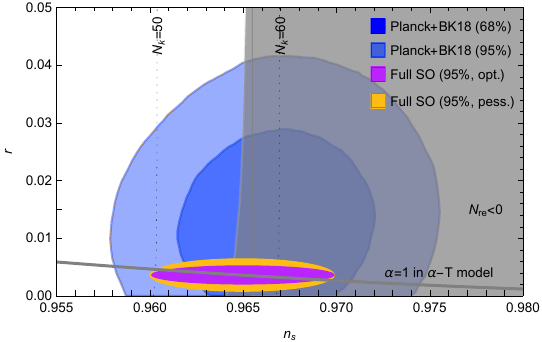}\\ \includegraphics[width=0.9\linewidth]{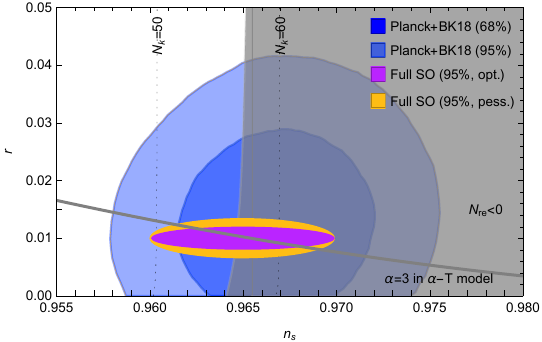}
	\caption{
    \emph{Upper panel:} 95\% confidence regions of the likelihood function \eqref{Eq:Likelihood} for benchmarks \ref{it:pessA} (yellow) and
    \ref{it:optA} (violet)
    compared to and current observational constraints from Planck+BK (blue).
    In the shaded gray area $\Nreh<0$ in the $\alpha$-T model \eqref{alpha V}; the dashed curves indicate specific values of $\Nk$, i.e., the number of $e$-folds between the horizon crossing of mode $k$ and the end of inflation. 
  The gray line indicates a curve of constant  $\alpha=1$ defined by \eqref{AlphaInAlphaT}, along this line $\Treh$ changes monotonically, with larger values towards the right. 
    \emph{Lower panel:} Same for $\alpha=3$ and benchmarks \ref{it:pessB} (yellow) and
    \ref{it:optB} (violet).
    } 
	\label{fig:ellipsesalphaT}
\end{figure}

\begin{figure}
	\centering
    \includegraphics[width=0.9\linewidth]{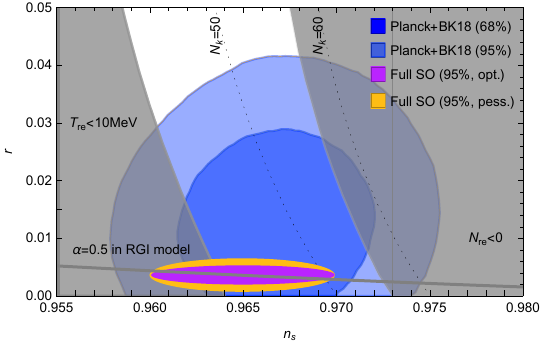}\\ \includegraphics[width=0.9\linewidth]{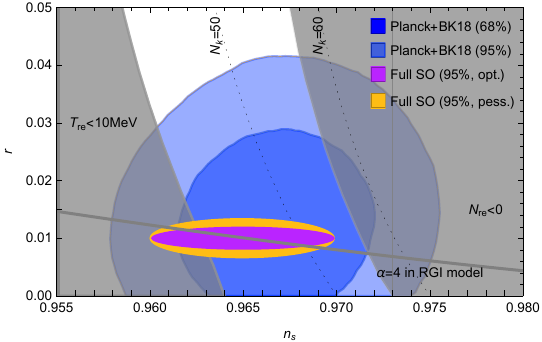}
	\caption{Same as Fig.~\ref{fig:ellipsesalphaT}, but for the RGI model \eqref{RGI V}. In the gray area on the left $\Treh < 10$ MeV, implying that the universe is not reheated sufficiently to assure successful BBN. 
     } 
	\label{fig:ellipsesRGI}
\end{figure}

\begin{figure}
	\centering
\includegraphics[width=0.9\linewidth]{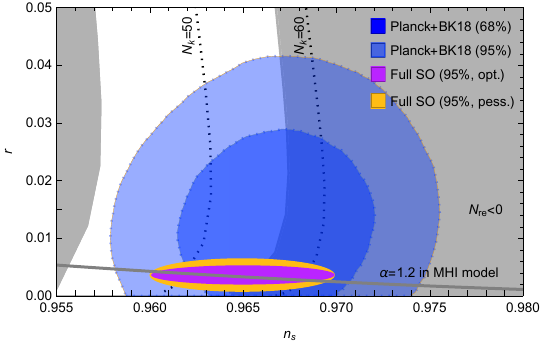}\\ \includegraphics[width=0.9\linewidth]{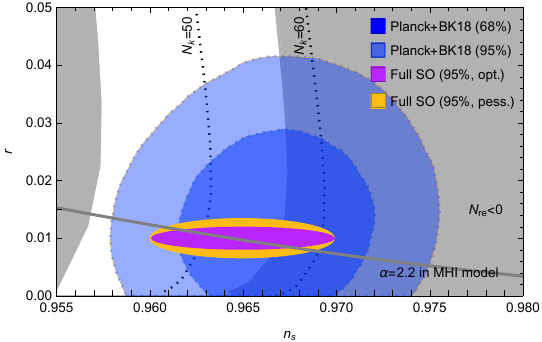}
	\caption{Same as Fig.~\ref{fig:ellipsesalphaT}, but for the MHI model \eqref{MHI V}. In the gray area on the left $\Treh < 10$ MeV, implying that the universe is not reheated sufficiently to assure successful BBN. 
     } 
	\label{fig:MHIellipses}
\end{figure}

In table \ref{PrincipalBenchmark} we consider several fiducial values $\nsbar$ and $\rbar$. 
For the case $\rbar = 0.0036$ and $\nsbar=0.965$ we obtain the values of $\sigmans$ and $\sigmar$ from Fig.~4 in \cite{SimonsObservatory:2025avm}, which followed the approach outlined in \cite{SimonsObservatory:2018koc} and \cite{Wolz:2023lzb}. 
These forecasts are based on the cumulative sensitivity after two years of operation with the original three SO SATs and eight years of operation with the full array of six SATs \footnote{
More precisely, the authors of \cite{SimonsObservatory:2025avm} assumed that the  first additional MF SAT starts observations half-way through the second year, and the final two instruments (a second additional MF SAT, and the additional LF SAT) start observations at the
start of the third year.}, 
and assuming 70\% delensing.
The sensitivities have been computed for two noise models, labelled \emph{pessimistic} and \emph{optimistic} in \cite{SimonsObservatory:2018koc}. 
Since the sensitivities to $n_s$ and $r$ are in first approximation independent of each other, we apply the same value of $\sigmar$ for all choices of $\nsbar$ in table \ref{PrincipalBenchmark}. 
We estimate the functional dependence of $\sigmar$ on $\rbar$ based on table 5 in \cite{Wolz:2023lzb}, where it was found that $\sigmar$ increases by roughly 15-20\% between $\rbar=0$ and $\rbar=0.01$ \footnote{We note that the interpretation of $\sigmar$ as a variance is only indicative; especially for $\rbar = 0.0036$ it is restricted to the $1\sigma$-level due to the non-Gaussian shape of the full likelihood for small values of $r$, while for larger values of $\rbar$ its range of validity increases. 
In any case, the ability to remove foregrounds will have a stronger impact on the precision at which $r$ can be measured than the dependence of $\sigmar$ on $\rbar$ studied in \cite{Wolz:2023lzb}, hence our estimates are sufficient for the present purpose.}. 
We compare the likelihood function to current observations and model predictions in Fig.~\ref{fig:ellipsesWI} 
and Figs.~\ref{fig:ellipsesalphaT}--\ref{fig:MHIellipses}.
The benchmarks labelled A and B are inspired by the the combined Planck 2018 and BICEP/Keck (Planck+BK) data \cite{BICEP:2021xfz}; 
benchmarks C and D are closer to the value of $n_s$ reported by the Atacama Cosmology Telescope \cite{AtacamaCosmologyTelescope:2025blo}.

\begin{table}
\centering
\begin{tabular}{c | c | c | c | c | c}
benchmark & $A_s$ & $\nsbar$ & $\sigmans$ & $\rbar$ & $\sigmar$\\
\hline\hline
Planck+BK & $10^{-10}e^{3.043}$ &  0.967  & 0.0036 & (0.01) &  (0.012) \\
\hline
\begin{enumerate*}[SO(pess.)A]
\item \label{it:pessA}
\end{enumerate*}
& $10^{-10}e^{3.043}$ &  0.965  & 0.0020 & 0.0036 & 0.0012  \\
\begin{enumerate*}[SO(pess.)B]
\item \label{it:pessB}
\end{enumerate*} & $10^{-10}e^{3.043}$ & 0.965  & 0.0020 & 0.01 & 0.0014 \\
\begin{enumerate*}[SO(pess.)C]
\item \label{it:pessC}
\end{enumerate*} & $10^{-10}e^{3.043}$ & 0.971  & 0.0020 & 0.0036 & 0.0012 \\
\begin{enumerate*}[SO(pess.)D]
\item \label{it:pessD}
\end{enumerate*} & $10^{-10}e^{3.043}$ & 0.974  & 0.0020 & 0.0036 & 0.0012 \\
\hline
\begin{enumerate*}[SO(opt.)A]
\item \label{it:optA}
\end{enumerate*} & $10^{-10}e^{3.043}$ & 0.965  & 0.0020 & 0.0036 & 0.0007\\
\begin{enumerate*}[SO(opt.)B]
\item \label{it:optB}
\end{enumerate*} & $10^{-10}e^{3.043}$ & 0.965  & 0.0020 & 0.01 & 0.0008 \\
\begin{enumerate*}[SO(opt.)C]
\item \label{it:optC}
\end{enumerate*} & $10^{-10}e^{3.043}$ & 0.971  & 0.0020 & 0.0036 & 0.0007\\
\begin{enumerate*}[SO(opt.)D]
\item \label{it:optD}
\end{enumerate*} & $10^{-10}e^{3.043}$ & 0.974  & 0.0020 & 0.0036 & 0.0007 
\end{tabular}
\caption{Fiducial values and uncertainties that define our observational benchmark scenarios, compared to the parameters fitted to the Planck+BK results, obtained from Fig.~5 in \cite{BICEP:2021xfz}. The error on $A_s$ can be neglected compared to the other uncertainties \cite{Drewes:2023bbs}, hence we fix $A_s$ to its best fit value from \cite{Planck:2018vyg}.
}
\label{PrincipalBenchmark}
\end{table}

\section{Results}

We find that observations with  SO in the benchmark scenarios given in tabel \ref{PrincipalBenchmark} can dramatically improve our knowledge on  the reheating temperature $\Treh$ and the underlying microphysical parameters in all models under consideration.

\subsection{QCD-driven Warm Inflation}

\begin{figure}
	\centering
\includegraphics[width=0.9\linewidth]{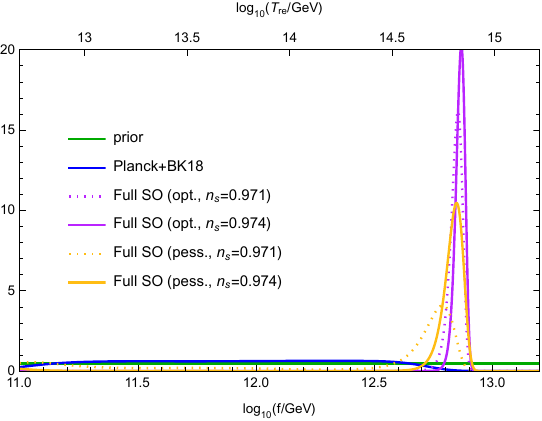}
	\caption{One-dimensional posterior distribution function for the QCD-WI model \eqref{eq:lag} in benchmark scenarios \ref{it:pessC}, \ref{it:pessD}, \ref{it:optC} and \ref{it:optD}, as indicated in the plot. The blue line represents the posterior from Planck+BK data.
     } 
	\label{fig:WIposterior}
\end{figure}

The resulting posterior for $\x$  is shown in Fig.~\ref{fig:WIposterior}.
Since the shape of this posterior considerably deviates from a Gaussian, it is instructive to compare different characteristics of the distribution.
In  table \ref{WIparameters} we list the variances as well as the maximum a posteriori (MAP) estimate, the full width at half maximum (FWHM) and peak mass. 
Approximating $\frac{\pi^2}{30} g_{*} \Treh^4  \approx \frac{\GG}{4H} (\V'/(3H + \GG))^2$ \footnote{Practically this amounts to $\Nreh=0$. In warm inflation scenarios there is no reheating in the literal sense because the thermal plasma of particles is already present during inflation. However, in principle a number of $e$-folds $\Nreh$ could pass between the moment when $\ddot{a}<0$ and the moment when the radiation's energy density exceeds that of $\varphi$. In \cite{Berghaus:2025dqi} it was estimated that this period is very brief; a more detailed investigation would require computing $\GG$ throughout the entire cosmic history, which goes beyond the scope of the present work.  
}
this directly translates a given value of $\x$ into a reheating temperature $\Treh$.
The results show that SO could, for the benchmarks considered here, measure the order of magnitude of the decay constant $f$ and the corresponding reheating temperature $\Treh$ even in the pessimistic scenarios. 
In the optimistic a measurement at the level of a few percent could be possible at the $1 \sigma $-level. The $99.7\%$ highest posterior density (HPD) intervals in $\log_{10}\frac{\Treh}{\rm GeV}$ are $[12.72, 12.91]$ for \ref{it:optC} and $[12.78, 12.91]$ for \ref{it:optD}, respectively.

\begin{table}
\centering
\begin{tabular}{c | c c c c c}
benchmark  & $\x$ & MAP$_\x$ & FWHM$_\x$ & mass & $\log_{10}\frac{\Treh}{\rm GeV}$ \\
\hline
\ref{it:pessC} & $12.36\pm 0.62$ & 12.79 & 0.15 & 49\% & $14.17\pm0.60$\\
\ref{it:optC}  & $12.84\pm0.04$ & 12.85 & 0.06 & 73\% & $14.56\pm0.03$\\
\ref{it:pessD} & $12.81\pm0.21$ & 12.85 & 0.09 & 72\% & $14.53\pm0.20$\\
\ref{it:optD}  & $12.86\pm0.02$ & 12.87 & 0.05 & 74\% & $14.57\pm0.01$\\
\end{tabular}
\caption{Parameters characterizing the posteriors displayed in Fig.~\ref{fig:WIposterior}
for the warm inflation model \eqref{eq:lag}.
\label{WIparameters}}
\end{table}

%%%%%%%%%%%%%%%%%%%%%%%%%%%%%%%%%%%%%%%%%%%%%%
%%%%%%%%%%%%%%%%%%%%%%%%%%%%%%%%%%%%%%%%%%%%%%%
\subsection{Standard (cold) plateau inflation}
\begin{figure}
	\centering
\includegraphics[width=0.9\linewidth]{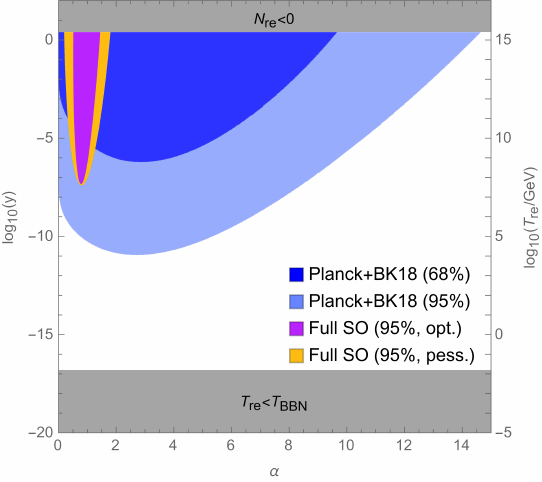}\\ \includegraphics[width=0.9\linewidth]{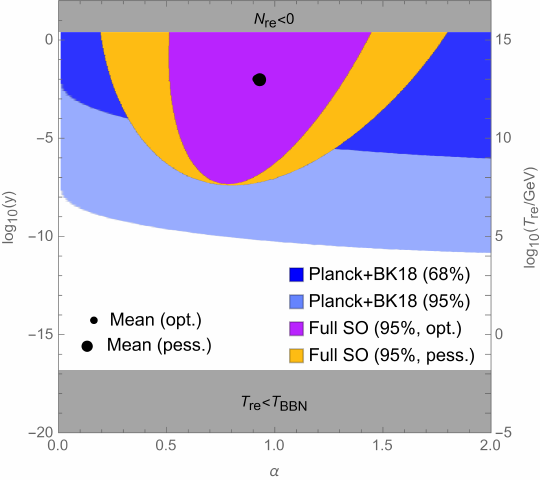}
\includegraphics[width=0.8\linewidth]{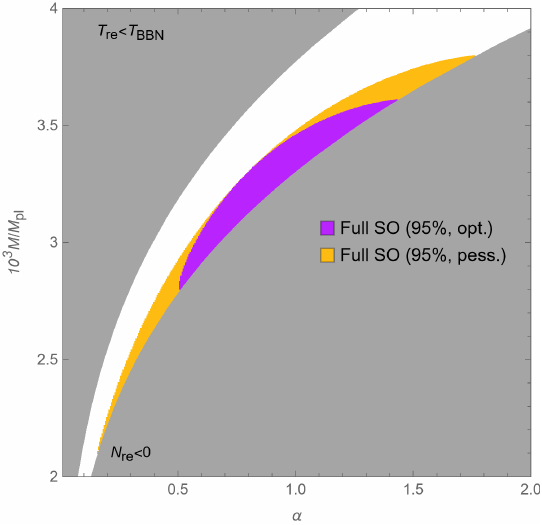}
	\caption{\emph{Upper panel:} Two-dimensional posterior distribution
    for $\alpha$ and $\Treh$ in the $\alpha$-T model \eqref{alpha V} from Planck+BK (blue), compared for forecasts for the benchmark scenarios \ref{it:pessA} (yellow) and \ref{it:optA} (violet) with $\rbar=0.0036$. The upper and lower gray regions are excluded by the requirements $\Nreh > 0$ and $\Treh > T_{\rm BBN}$. The black dots mark the best fit in both scenarios, they coincide in this case.
    \emph{Middle panel:} Zoom into the region preferred in scenarios \ref{it:pessA} and \ref{it:optA}. 
    \emph{Lower panel:} Corresponding two-dimensional posterior distributions in the $\M$-$\alpha$-plane.
     } 
	\label{fig:alphaT2D}
\end{figure}

\begin{figure}
	\centering
\includegraphics[width=0.9\linewidth]{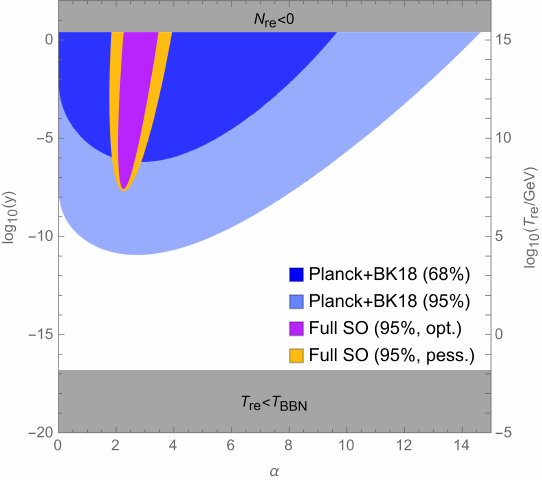}\\ \includegraphics[width=0.9\linewidth]{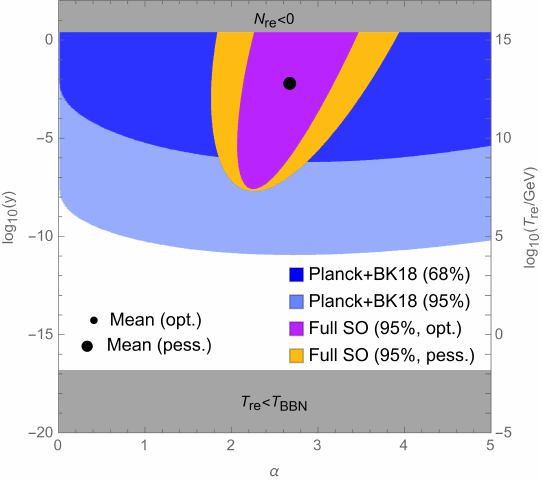}
\includegraphics[width=0.9\linewidth]{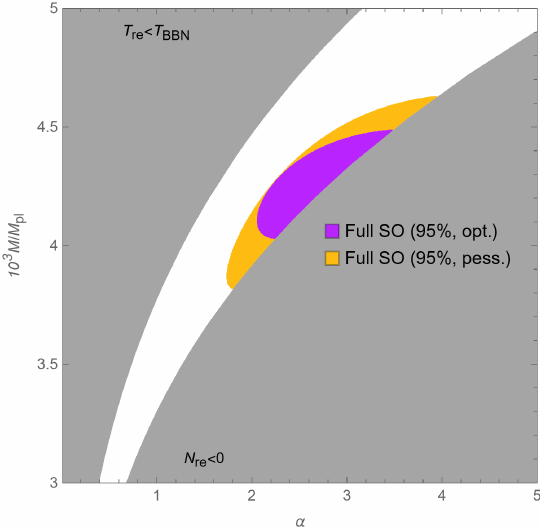}
	\caption{
    Same as Fig.~\ref{fig:alphaT2D}, but for the scenarios \ref{it:pessB} (yellow) and \ref{it:optB} (violet) with $\rbar=0.01$.
     } 
	\label{fig:alphaT2Dr001}
\end{figure}

%\ref{it:pessA}
An important implication of measuring $r$ in the plateau models
\eqref{alpha V}-\eqref{MHI V}
is that it would determine the scale of inflation
$\M \sim \mpl(
     3\pi^2
     A_s r / 2
     )^{1/4}$, where the symbol $\sim$ indicates that we have omitted an $\alpha$-dependent prefactor \footnote{The precise relations between $\M$, $\alpha$ and observables (as well as the quantities $\varphi_k$, $N_k$, $\Nreh$ etc.) used below are well-known for the models \eqref{alpha V}-\eqref{MHI V}; for a summary we refer the reader to \cite{Drewes:2023bbs,Drewes:2026uor}.}.
It is instructive to assess how well SO observations will be able to simultaneously measure $\M$ and $\alpha$ if the uncertainty due to reheating is fully taken into account. We find that this SO observations can pin down both parameters up to factors of order one in all benchmarks considered here (see Figs.~\ref{fig:alphaT2D}, \ref{fig:alphaT2Dr001}, \ref{fig:RGI2D}, \ref{fig:RGI2Dr001}, \ref{fig:MHI2D}, \ref{fig:MHI2Dr001}).
      Information on $\M$ and $\alpha$ translates into knowledge on
$m_\phi 
\approx \sqrt{24 \pi^2 A_s} \mpl\Nk$ \cite{Drewes:2026uor}
and all inflaton self-interactions by means of the expansion \eqref{Taylor}.

When it comes to constraining reheating itself, 
in the plateau models \eqref{alpha V}-\eqref{MHI V}
a measurement of $r$ is less constraining on $\Treh$ 
than for QCD-WI for two reasons.
Firstly, these models contain two parameters $\{\M,\alpha\}$ in the potential, implying that the constraint that a given measurement imposes on each of them tends to be weaker on general grounds.
As a result, a measurement of $\Treh$ (or likewise $\x$) is practically only possible if one of the parameters is fixed from other considerations, as already noted in \cite{Drewes:2023bbs}.
Secondly, the dependence of 
$\Treh$ on $r$ is weaker. 
This can be seen by noticing that each choice of $\alpha$ defines a line in the $n_s$-$r$-plane  along which $\Treh$ changes, with larger values of $\Treh$ corresponding to larger $n_s$ and smaller $r$.
For small $r$ this gray line is almost horizontal in Figs.~\ref{fig:ellipsesalphaT}-\ref{fig:MHIellipses}, while the corresponding black line in Fig.~\ref{fig:ellipsesWI} is almost vertical. As a result of this,  sensitivity of SO to reheating is not owed to its ability to detect primordial gravitational waves alone, but also benefits from the reduced $\sigmans$ compared to Planck+BK.
Further improvement can be achieved with data from optical surveys and 21cm tomography, in particular with the EUCLID satellite and the Square Kilometer Array \cite{Sprenger:2018tdb}.

\begin{table}
	\centering
    \begin{tabular}{c|c|c|c|c|c}
		 model & $\alpha$ & benchmark & $\x$ & 
           $10^3 \M/\mpl$
         & ${\rm log}_{10}\frac{T_{\text{re}}}{\text{GeV}}$\\
		%\hline
		\hline
        \hline
      $\alpha$-T & 1 & \ref{it:pessA} &  $-2.0\pm1.9$ & $3.36\pm0.05$ & $12.8\pm1.9$ \\
      $\alpha$-T & 1 & \ref{it:optA} & $-1.8\pm1.7$ & $3.35\pm0.04$ & $13.1\pm1.7$ \\
      $\alpha$-T & 3 & \ref{it:pessB} &  $-1.6\pm1.6$ & $4.39\pm0.05$ & $13.2\pm1.6$ \\
      $\alpha$-T & 3 & \ref{it:optB} & $-1.0\pm1.1$ & $4.37\pm0.03$ & $13.8\pm1.1$ \\
        \hline
		RGI & 0.5 & \ref{it:pessA} & $-13.2\pm2.6$ & $3.24\pm0.05$ & $1.9\pm2.6$ \\
		%\hline
        RGI & 0.5 & \ref{it:optA} &  $-13.4\pm2.4$ & $3.24\pm0.05$ & $1.7\pm2.4$ \\
       % \hline
		 RGI & 4 & \ref{it:pessB} &  $-12.4\pm2.7$ & $4.30\pm0.07$ & $2.6\pm2.6$ \\
       % \hline
      RGI & 4 & \ref{it:optB} & $-12.7\pm2.2$ & $4.31\pm0.06$ & $2.3\pm2.2$ \\
      \hline
      MHI & 1.2 & \ref{it:pessA} &  $-3.5\pm2.4$ & $3.25\pm0.05$ & $11.4\pm2.4$ \\
      MHI & 1.2 & \ref{it:optA} & $-3.4\pm2.3$ & $3.25\pm0.05$ & $11.5\pm2.3$ \\
      MHI & 2.2 & \ref{it:pessB} &  $-4.4\pm2.5$ & $4.27\pm0.07$ & $10.4\pm2.5$ \\
      MHI & 2.2 & \ref{it:optB} & $-4.3\pm2.2$ & $4.27\pm0.06$ & $10.6\pm2.2$
	\end{tabular}
	\caption{One-dimensional posterior mean values and variances for $\M$, $\x$ and $\Treh$ in the three theoretical models \eqref{alpha V}, \eqref{RGI V} and \eqref{MHI V} for the observational benchmarks defined in Table \ref{PrincipalBenchmark}.    }
	\label{ErrorBars}
\end{table}

\begin{figure}
	\centering
\includegraphics[width=0.9\linewidth]{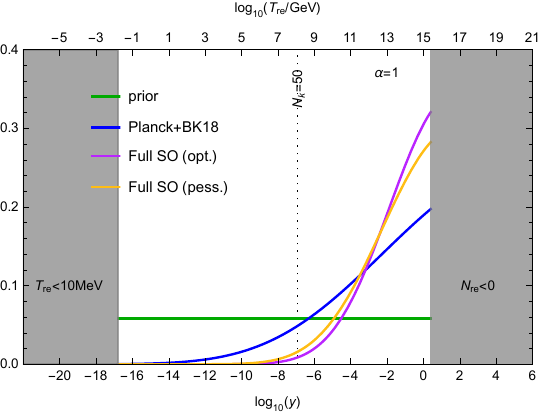}\\
\includegraphics[width=0.9\linewidth]{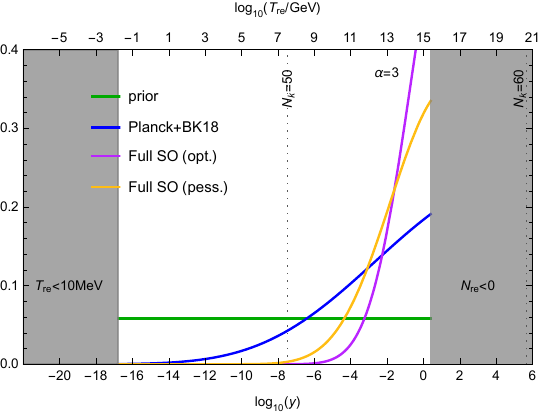}
	\caption{\emph{Upper panel:} One-dimensional posteriors for the $\alpha$-T model \eqref{alpha V} 
    with fixed $\alpha=1$ in scenarios  \ref{it:pessA} (yellow) and \ref{it:optA} (violet), compared to the posterior obtained from Planck+BK (blue) and the prior (green).
    \emph{Lower panel:} Same for $\alpha=3$ and scenarios \ref{it:pessB} (yellow) and \ref{it:optB} (violet).} 
	\label{fig:alphaT1D}
\end{figure}

\begin{figure}
	\centering
\includegraphics[width=0.9\linewidth]{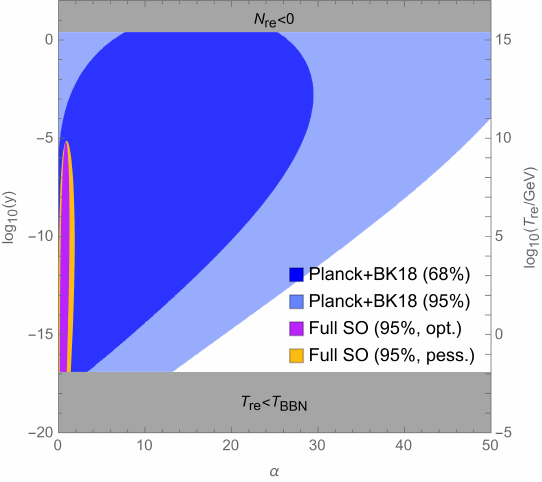}\\ 
\includegraphics[width=0.9\linewidth]{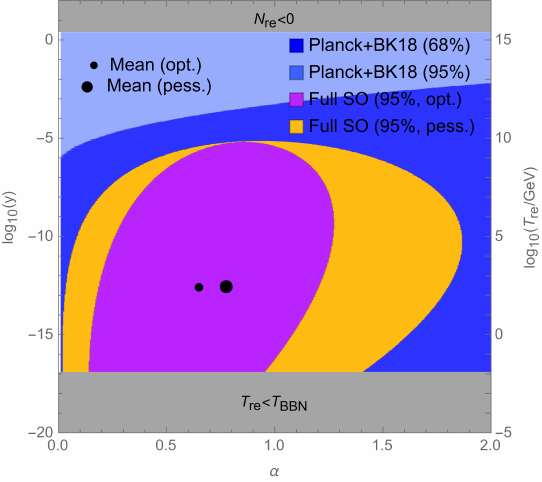}
\includegraphics[width=0.9\linewidth]{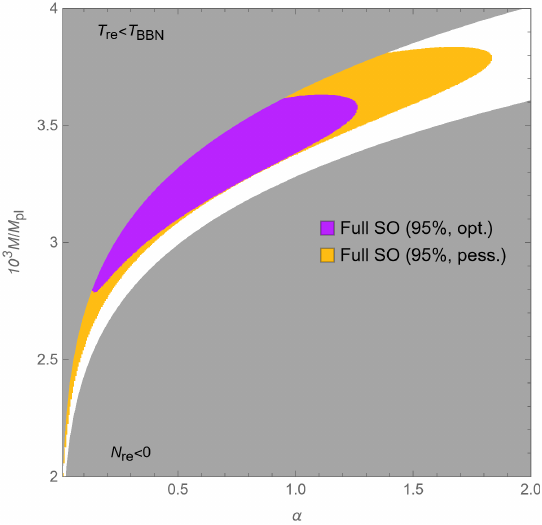}
	\caption{Same as Fig.~\ref{fig:alphaT2D}, but for the RGI model \eqref{RGI V} in  scenarios \ref{it:pessA} (yellow) and \ref{it:optA} (violet).
     } 
	\label{fig:RGI2D}
\end{figure}

\begin{figure}
	\centering
\includegraphics[width=0.9\linewidth]{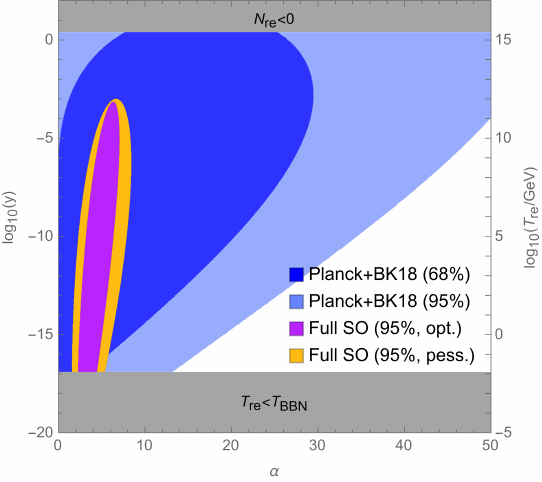}\\ 
\includegraphics[width=0.9\linewidth]{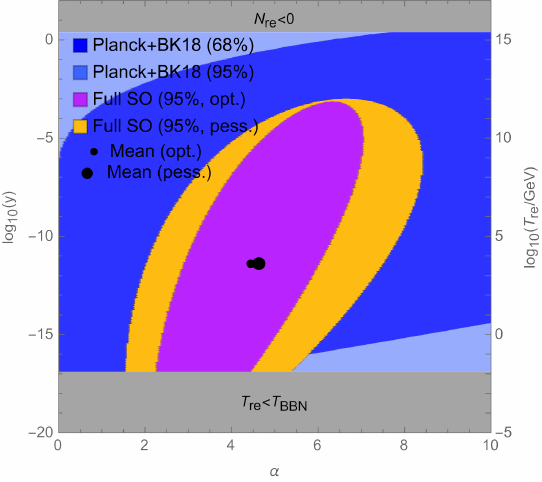}
\includegraphics[width=0.9\linewidth]{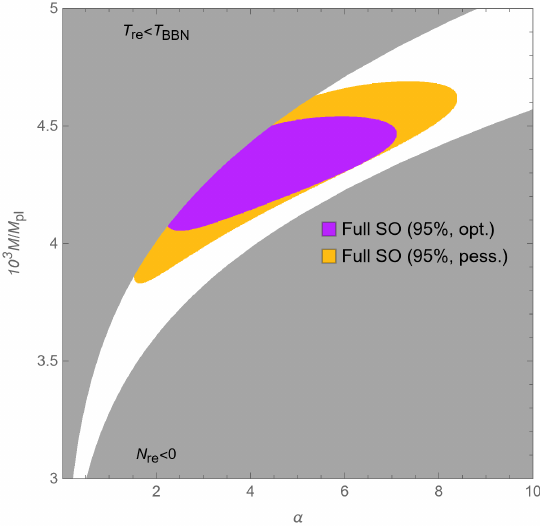}
	\caption{%RGI 0.01
     Same as Fig.~\ref{fig:alphaT2Dr001}, but for the RGI model \eqref{RGI V} in scenarios  \ref{it:pessB} (yellow) and \ref{it:optB} (violet).
     } 
	\label{fig:RGI2Dr001}
\end{figure}

\paragraph{$\alpha$-attractor T-model \eqref{alpha V}:}
%%%%%%%%%%%%%%%%%%%%%%%%%%%%%%%%%%%%%%%%%%%%%%%%%%%%%%%%%%%%%%%%%%%%%%%%%%%%%%%%%%%%%%
Fig.~\ref{fig:ellipsesalphaT} shows that a sizeable fraction of the model's parameter space is already excluded by current Planck+BK data. 
Recalling that the reheating temperature increases from left to right along each line defined by
\begin{equation}\label{AlphaInAlphaT}
    \alpha=\frac{4r}{3(1-n_s)(4(1-n_s)-r)},
\end{equation}
it is evident that Planck+BK imposes a rather strong lower bound on $\Treh$.
In Figs.~\ref{fig:alphaT2D}
and
\ref{fig:alphaT2Dr001}
we show the two-dimensional posteriors after SO observations for the fiducial values given in table \ref{PrincipalBenchmark}.
SO can simultaneously measure the parameters $\M$ and $\alpha$ in spite of the uncertainty from reheating. 
For the favoured values of $\alpha$, the lower bound on $\Treh$ 
at the 95\% confidence level (CL) tightens by 2-3 orders of magnitude, but the allowed range still covers several orders of magnitude. 
In Fig.~\ref{fig:alphaT1D} we show one-dimensional posteriors in $\Treh$ (or likewise $y$) for fixed $\alpha$. Basic properties of these posteriors are summarised in table \ref{ErrorBars}. The physical interpretation of the variances has to be treated with care, as the posteriors are highly non-Gaussian and skewed. However, they may still be used in an indicative way and demonstrate that SO could measure the order of magnitude of $\Treh$ and $y$ for given $\alpha$. 
Since \eqref{PerturbativityContraintsPleteau} is violated in most of the allowed parameter region, a 
reconstruction of the thermal history during reheating is likely to be hampered by strong feedback effects \cite{Drewes:2026uor}.

\begin{figure}
	\centering
\includegraphics[width=0.9\linewidth]{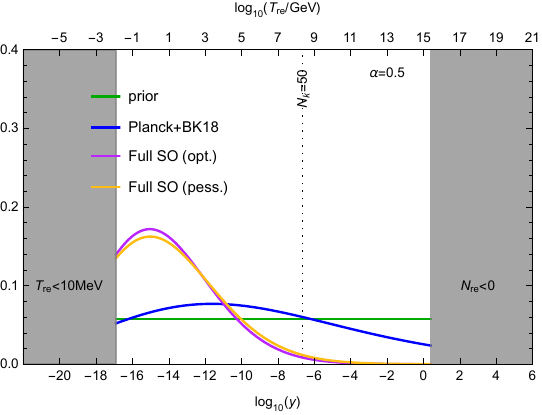}\\
\includegraphics[width=0.9\linewidth]{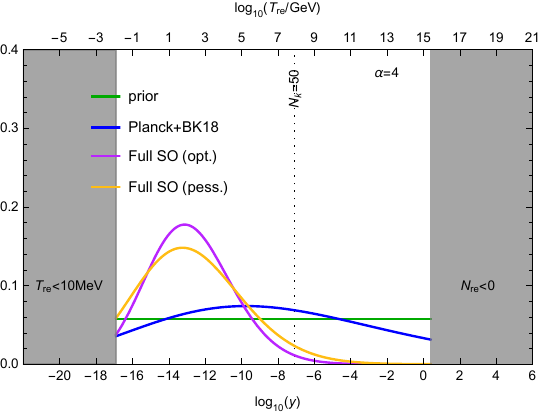}
	\caption{\emph{Upper panel:} Same as Fig.~\ref{fig:alphaT1D}, but for the RGI model \eqref{RGI V} with $\alpha=1/2$ in scenarios \ref{it:pessA} (yellow) and \ref{it:optA} (violet). \emph{Lower panel:} Same for $\alpha=4$ and in scenarios \ref{it:pessB} (yellow) and \ref{it:optB} (violet).}  
	\label{fig:RGI1D}
\end{figure}

\begin{figure}
	\centering
\includegraphics[width=0.9\linewidth]{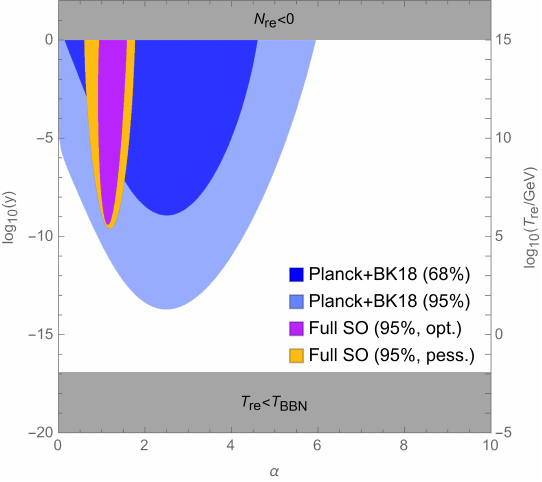}\\ 
\includegraphics[width=0.9\linewidth]{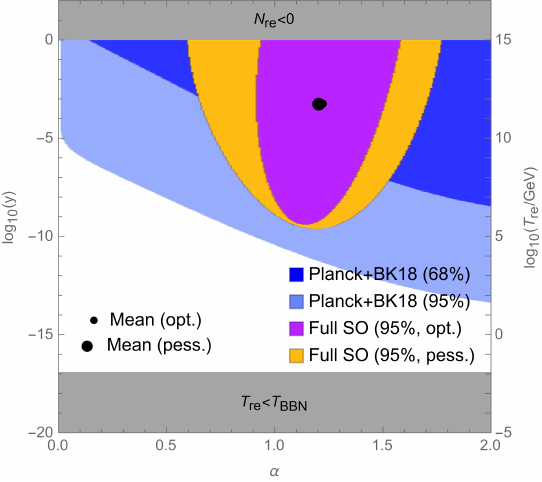}
\includegraphics[width=0.9\linewidth]{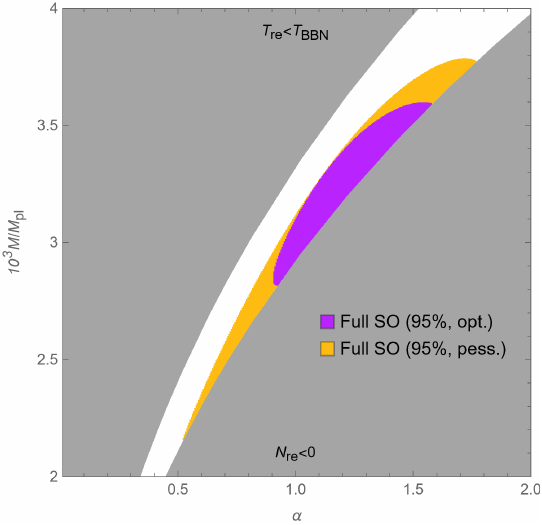}
	\caption{Same as Fig.~\ref{fig:alphaT2D}, but for the MHI model \eqref{MHI V} in scenarios scenarios \ref{it:pessA} (yellow) and \ref{it:optA} (violet).
     } 
	\label{fig:MHI2D}
\end{figure}

\begin{figure}
	\centering
\includegraphics[width=0.9\linewidth]{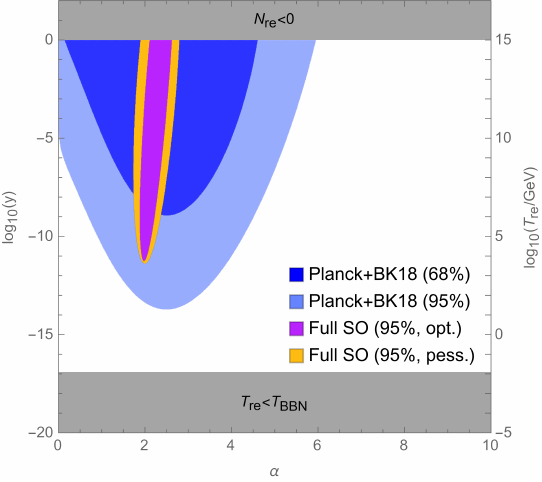}\\ 
\includegraphics[width=0.9\linewidth]{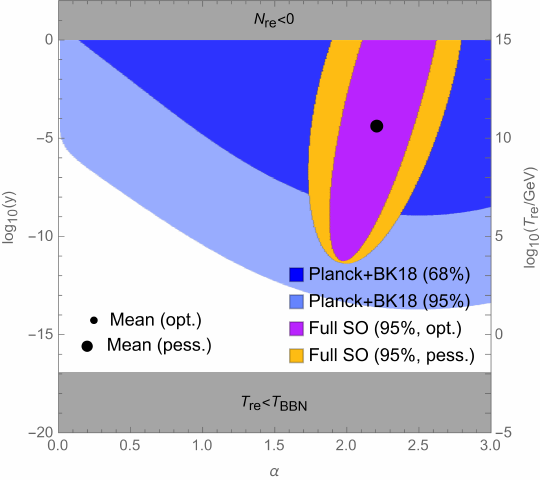}
\includegraphics[width=0.9\linewidth]{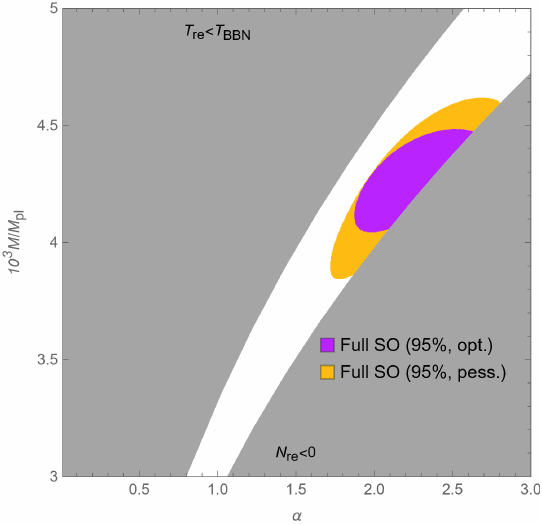}
	\caption{%MHI 0.01
     Same as Fig.~\ref{fig:alphaT2Dr001}, but for the MHI model \eqref{MHI V} in scenarios scenarios \ref{it:pessB} (yellow) and \ref{it:optB} (violet).
     } 
	\label{fig:MHI2Dr001}
\end{figure}

%%%%%%%%%%%%%%%%%%%%%%%%%%%%%%%%%%%%%%%%%%%%%%%%%%%%%%%%%%%%%%%%%%%%%%%%%%%%%
\paragraph{Radion gauge inflation \eqref{RGI V}:}
From Fig.~\ref{fig:ellipsesRGI} it is already clear that current Planck+BK data practically does not constrain $\Treh$ for a wide range of $\alpha$: 
For $r\lesssim 0.03$ the lines defined by 
\begin{equation}\label{珍珠奶茶}
    \alpha=\frac{432r^2}{(8(1-n_s)+r)^2(4(1-n_s)-r)}
\end{equation}
are cut on both ends by constraints that are unrelated to CMB observations: On the right the upper bound on $\Treh$ is imposed by the physical requirement $\Nreh>0$ (which is essentially energy conservation), on the left the lower limit is $T>T_{\rm BBN}$.
This is reflected in the posteriors  in Figs.~\ref{fig:RGI2D}
and
\ref{fig:RGI2Dr001}: 
For $\alpha \lesssim 10$ there is essentially no constraint on $\Treh$ at the 95\% CL.
The allowed range of $\alpha$ extends beyond $\alpha >50$, and for $\alpha >10$ there is a lower bound on $\log_{10}\Treh$ that approximately grows linearly with $\alpha$.
SO will change this situation dramatically. 
Observations in the benchmark considered here will not only be able to measure $\alpha$ accurately, but also impose an upper bound $\Treh < 10^{10}$ GeV. 
However, an actual measurement of $\Treh$  again requires fixing $\alpha$ from other considerations. In Fig.~\ref{fig:RGI1D} we display one-dimensional posteriors for two choices of $\alpha$, some of their properties are summarised in table \ref{ErrorBars}.
Within the $1 \sigma$-regions the posteriors are approximately Gaussian, so that the interpretation of the variances is straightforward.
Finally, we note that the condition \eqref{PerturbativityContraintsPleteau} is fulfilled in the parameter region preferred by the observational benchmark scenarios considered here, 
allowing for an unambiguous interpretation of the posteriors in terms of the microphysical inflaton coupling $y$, and potentially a reconstruction of the thermal history during reheating to predict the abundance of thermal relics~\cite{Drewes:2026uor}.

\begin{figure}
	\centering
\includegraphics[width=0.9\linewidth]{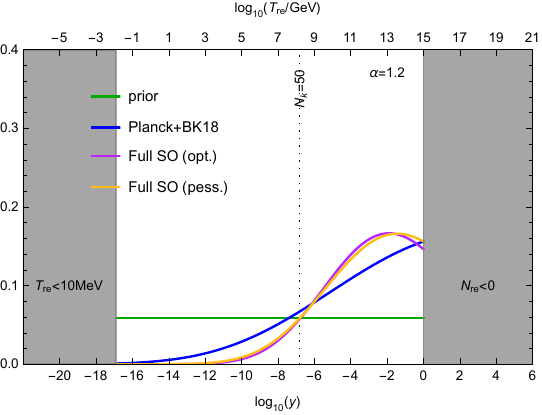}\\
\includegraphics[width=0.9\linewidth]{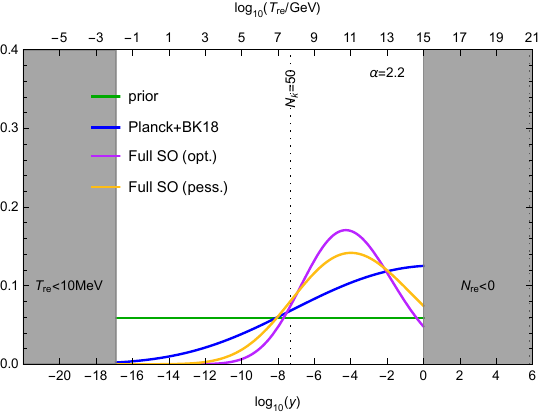}
	\caption{\emph{Upper panel:} Same as Fig.~\ref{fig:alphaT1D}, but for the MHI model \eqref{MHI V} with $\alpha=1.2$ in scenarios \ref{it:pessA} (yellow) and \ref{it:optA} (violet). \emph{Lower panel:} Same for $\alpha=2.2$ and in scenarios \ref{it:pessB} (yellow) and \ref{it:optB} (violet).} 
	\label{fig:MHI1D}
\end{figure}

%%%%%%%%%%%%%%%%%%%%%%%%%%%%%%%%%%%%%%%%%%%%%%%%%%%%%%%%%%%%%%%%%%%%%%%%%%%
\paragraph{Mutated hilltop inflation \eqref{MHI V}:}
Again recalling that $\Treh$ is lower towards the left of Fig.~\ref{fig:MHIellipses} it is clear that current Planck+BK data already impose a lower bound on $\Treh$ and an upper bound on $\alpha$.
Figs.~\ref{fig:MHI2D} and \ref{fig:MHI2Dr001}
demonstrate how these constraints will improve with SO.
As in the $\alpha$-T and RGI models, $\M$ and $\alpha$ can be measured simultaneously.
The current lower bound on $\Treh$ can be expected to improve by almost two orders of magnitude. 
Fig.~\ref{fig:MHI1D} and table \ref{ErrorBars} demonstrate that 
a measurement of $\Treh$ is possible if $\alpha$ can be fixed from other considerations. 
A sizeable part of the posterior lies in the regime where \eqref{PerturbativityContraintsPleteau} is fulfilled.

\medskip

\begin{figure}
	\centering
\includegraphics[width=0.9\linewidth]{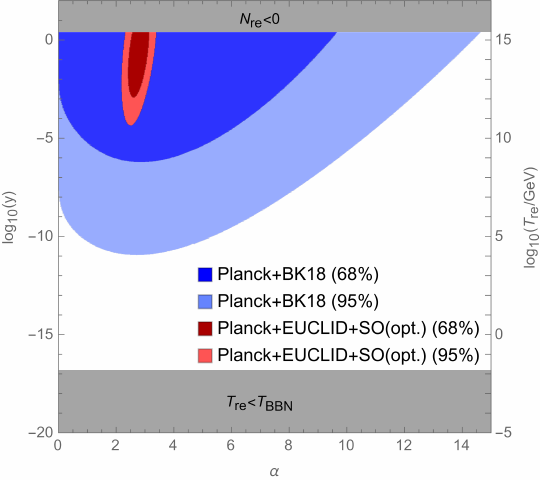}\\ 
\includegraphics[width=0.9\linewidth]{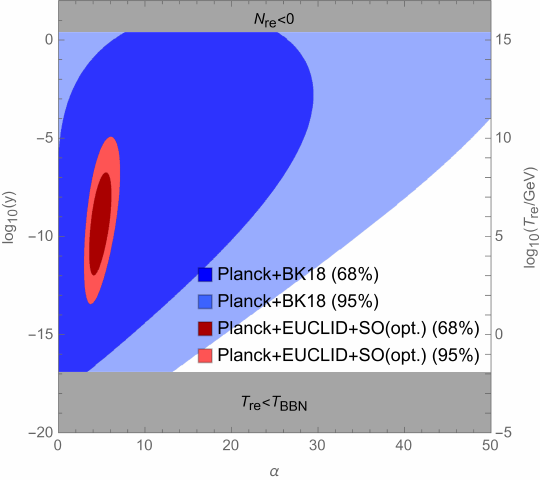}
\includegraphics[width=0.9\linewidth]{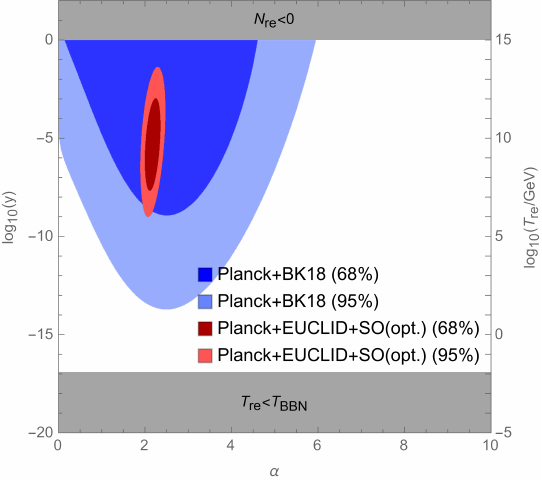}
	\caption{  
    Posteriors obtained by combining the optimistic SO estimate $\sigmar=0.0008$  with the forecast $\sigmans=0.00085$ for EUCLID \cite{Sprenger:2018tdb} in the $\alpha$-T model (upper panel), RGI model (middle panel) and MHI model (lower panel) in benchmark scenarios with $\rbar=0.01$ and $\nsbar=0.964$ (for MHI and $\alpha$-T) or $\nsbar=0.967$ (for RGI). A similar sensitivity could be achieved with the Square Kilometer Array \cite{Sprenger:2018tdb}.} 
	\label{EUCLID}
\end{figure}

Before concluding, we take a brief glimpse at the perspectives to probe reheating 
%could further improve 
if SO data is combined with other observations, in particular 
the Euclid satellite \cite{EUCLID:2011zbd} and the Square Kilometre Array \cite{Maartens:2015mra}, 
both of which are expected to considerably reduce the uncertainty of $n_s$ by mapping the distribution of matter in the universe \cite{Sprenger:2018tdb} and may even be able to detect its scale dependence (running). Combining data from different instruments is notoriously difficult, and the extraction of $n_s$  is further complicated by nonlinear structure formation, hence Fig.~\ref{EUCLID} should be regarded as a gaze into the crystal ball rather than a reliable forecast. 
It is nevertheless instructive to take soundings of what could possibly be achieved with existing technology.
Fig.~\ref{EUCLID} shows that the reduction in $\sigmans$ would for the first time enable an actual measurement of $\Treh$ without fixing $\alpha$ (in the sense of imposing both an upper and a lower bound from data rather than the requirements included in the prior), which cannot be achieved by SO alone. %If nothing else, 
This simple estimate demonstrates that measuring $n_s$ more accurately can be a game-changer rather than an incremental improvement in the present context.

\section{Discussion and conclusion}

We studied the perspectives to constrain the reheating temperature $\Treh$ and the microphysical inflaton coupling with the expanded SO. Our analysis is based on observational benchmark scenarios in which SO discovers primordial gravitational waves with fiducial values $\rbar= 0.0036$ or
$\rbar = 0.01$ for the tensor-to-scalar ratio. In addition, SO's ability to probe reheating also benefits from its improved sensitivity to the spectral index $n_s$ compared to Planck+BK.

For QCD-WI, $\Treh$ and the inflaton coupling to gluons can both be determined with an error-bar at the percent level. Such a measurement would imply a clear prediction for inflaton searches in axion experiments on Earth. This opens up a unique opportunity to probe the mechanism that set the initial conditions of the hot big bang. 

In the three conventional (cold) plateau models of inflation that we considered -- $\alpha$-attractors, RGI and MHI -- SO can simultaneously measure $\M$ and $\alpha$, and tighten the bound on $\Treh$ by orders of magnitude. If $\alpha$ can be fixed from other considerations, then SO could also measure the order of magnitude of $\Treh$. In the RGI and MHI models such a measurement could be translated into information on the microphysical inflaton coupling to other fields, a necessary condition to reconstruct the thermal history during reheating; in the $\alpha$-T model such a reconstruction is likely to be hampered by strong feedback effects.

Our results demonstrate that the expanded SO is a versatile and powerful tool to probe the origin of our Universe and the connection to theories of particle physics. 
While SO alone can already improve our current level of knowledge by orders of magnitude, combining SO with other probes -- space-born CMB observations, optical surveys, 21cm tomography and gravitational wave detectors -- will  open a new era in cosmology in which the physics of reheating can be probed for the first time.

\section*{Acknowledgements}
MaD would like to thank
Lyman Page for the discussions during his visit at UCLouvain and David Alonso for his feedback on the SO sensitivity to tensor perturbations. This work has been partially funded by the Deutsche Forschungsgemeinschaft (DFG, German Research Foundation) - SFB 1258 - 283604770. 

\begin{appendix}

\section{Relation between model parameters and CMB observables}\label{TrefromCMB}

The relation between CMB observables and model parameters in warm inflation is a matter of ongoing research.
In the present work we use the relations provided in \cite{Berghaus:2025dqi} to constrain the model \eqref{eq:lag}, which are based on the approach used in \cite{Mirbabayi:2022cbt}. Using the difference to other recent computations \cite{Ballesteros:2023dno,ORamos:2025uqs,Laine:2025rll} as a proxy for the theory uncertainty
\footnote{A possible way to systematically address those uncertainties could be the
Schwinger-Keldysh formalism (previously applied to the background evolution \cite{Buldgen:2019dus}) in combination with the open effective field method~\cite{Salcedo:2024smn}.}, 
we conclude that the theoretical error bar is currently still smaller than the observational one. In all approached, equations have to be solved numerically, and no analytic relation between $\{A_s,n_s,r\}$ and model parameters or $\Treh$ is known.

For  standard (cold) single field inflation, the connection between $\Treh$ and
$\{A_s,n_r,r\}$ can be expressed in terms of comparably simple analytic formulae \footnote{The relations in Apendix \ref{TrefromCMB} hold at leading order in the slow-roll parameters, see appendix B in \cite{Drewes:2023bbs} for a discussion of higher order corrections.}. 
 For completeness, we summarize the well-known relevant equations
we use to derive bounds on the models models \eqref{alpha V}, \eqref{RGI V} and \eqref{MHI V}
following \cite{Ueno:2016dim,Drewes:2017fmn}.
The number of $e$-folds $\Nk$ between 
the end of inflation and
the horizon-crossing of 
the wave number $k$  
is given by
\begin{equation}\label{Nk}
\Nk=
\ln\left(\frac{a_{\rm end}}{a_k}\right)=
\int_{\varphi_k}^{\varphi_{\rm end}}
\frac{H d\varphi}{\dot{\varphi}}
\approx
\frac{1}{\mpl^2}\int_{\varphi_{\rm end}}^{\varphi_k}d\varphi
\frac{\V }{\partial_\varphi \V }.
\end{equation}
Here $\varphi_k, H_k$, etc.~denote the values of  
$\varphi, H$, etc.~when $k$ crosses the horizon.

Assuming that the effective numbers of degrees of freedom contributing to the energy and entropy density are equal and constant during the relevant time, $g_\rho(T) = g_s(T) = g_\star$,
$\Nk$
is related to
$\Nreh$ by
\begin{equation}
	\label{Nre}
	\begin{split}
		N_{\rm re} &= \frac{4}{3\Bar{w}_{\rm re}-1}\Bigg[\Nk+\ln\left(\frac{k}{a_0 T_0}\right)+\frac{1}{4}\ln\left(\frac{40}{\pi^2g_*}\right)\\
		&\quad +\frac{1}{3}\ln\left(\frac{11g_{s*}}{43}\right)-\frac{1}{2}\ln\left(\frac{\pi^2M^2_{ pl}r A_s}{2\sqrt{\Vend}}\right)\Bigg],
	\end{split}
\end{equation}
Here $T_0=2.725~{\rm K}$ is the present temperature of the CMB, $a_0$ the present scale-factor, and $\wrehbar$ the averaged equation of state during reheating. 
Since the energy budget during reheating is (by definition) still dominated by the inflaton, $\wrehbar$ can in good approximation be computed from $\V$, leaving $\Nreh$ as the sole parameter that is not determined by the choice of $\V$.
Solving
\begin{equation}
	n_s=1-6\epsilon_k+2\eta_k~, \quad r=16\epsilon_k
	\label{nANDr}    
\end{equation}
fixes $\varphi_k$ in terms of 
$n_s$ and $r$.
During slow-roll ($\epsilon,\eta \ll 1$) the Hubble rate is
\begin{equation}
	\label{H_k}
	H^2_k=\frac{\V(\varphi_k)}{3\mpl^2}~
	=\pi^2 \mpl^2\frac{r A_s}{2}.
\end{equation}
 $\Vend$ and $\varphi_{\rm end}$ can be found by solving $\epsilon=1$ for $\varphi$, together with \eqref{nANDr} this yields
\begin{equation}
	\epsilon_k=\frac{r}{16}~,\quad \eta_k=\frac{n_s-1+3r/8}{2}.
\end{equation}
 $\Treh$ can be expressed in terms of the observables $\{A_s, n_s, r\}$,
by plugging \eqref{Nre} with \eqref{Nk} into \eqref{Tre}.
Using the definitions of the slow-roll parameters $\epsilon$ and $\eta$ this gives
\begin{equation}\label{TakaTukaUltras}
\frac{\partial_\varphi \V}{\V}\Bigl|_{\varphi_k}=\sqrt{\frac{r}{8\mpl^2}} \ , \quad
\frac{\partial^2_\varphi \V}{\V}\Bigl|_{\varphi_k}=\frac{n_s-1+3r/8}{2\mpl^2}.
\end{equation}
Thus, the three equations in \eqref{TakaTukaUltras} and \eqref{H_k} relate the potential $\V$ and its derivatives to $\{A_s,n_s,r\}$, allowing to express the quantities $\wrehbar$ and $\Nreh$ in in \eqref{Nre} in terms of observables.
This permits to compute the reheating temperature from the redshift relation
 \begin{equation}\label{Tre}
	\Treh=\exp\left[-\frac{3(1+\wrehbar)}{4}N_{\rm re}\right]\left(\frac{40 \ \Vend }{g_\star\pi^2}\right)^{1/4}\ .
\end{equation}
Since the dependence on $g_*$ is very weak, we can choose to use $\Treh$ instead of $\Nreh$ to parametrize the impact of reheating on the CMB.
Using the fact that reheating ends when $\GG=H$, 
this can be translated into 
\begin{equation}
	\GG|_{\GG=H}
	\approx \frac{\Treh^2}{\mpl}\frac{\sqrt{g_\star}}{3}\ .\label{GammaConstraint}
\end{equation}
When $\GG$ is calculable in terms of microphysical parameters, \eqref{GammaConstraint} permits to translate a constraint on $\Treh$ into information about these parameters, and hence the connection between inflation and particle physics.

\end{appendix}

\bibliography{bib.bib}

\end{document}